# Towards a compact transportable optical clock based on the octupole transition in $^{171}$Yb$^+$

Xuanjian Wang[1,2], Jian Cao[1,✉], Hualin Shu[1], Yi Yuan[1], Zehao Li[1,2], Pengcheng Fang[1], Qunfeng Chen[1], Xueren Huang[1,3,✉]

[1] Key Laboratory of Time Reference and Applications, Innovation Academy for Precision Measurement Science and Technology, Chinese Academy of Sciences, Wuhan 430071, China

[2] University of Chinese Academy of Sciences, Beijing 100049, China

[3] Wuhan Institute of Quantum Technology, Wuhan 430206, China

✉ e-mail:caojian@apm.ac.cn, hxueren@apm.ac.cn

**Abstract**: Optical clocks have extremely attractive applications in many fields, including time-frequency metrology, validation of fundamental physical principles, and relativistic geodesy. The 467 nm octupole transition in $^{171}$Yb$^+$ ion exhibits intrinsic insensitivity to magnetic field and an ultra-long clock state lifetime of 1.6 years. In addition, the entire laser system can be realized by semiconductor technologies, rendering this platform uniquely advantageous for developing high-precision, compact and transportable optical clocks. Here, we report the development of a compact optical clock based on the 467 nm transition of a single $^{171}$Yb$^+$ ion. Using a narrow linewidth 467 nm laser to interrogate the clock transition, we obtain a near-Fourier-limited linewidth of 2.3 Hz in an integrated ion trapping system. Self-comparison demonstrated a frequency instability of $2.2\times10^{-15}/\sqrt{\tau/s}$ with an interrogation time of 180 ms, which reaches the high parts in $10^{-18}$ level with an averaging time of only one day. These work laid the technical foundation for the subsequent clock systematic evaluation and the packaging of each subsystem into an engineering prototype with high-precision at the level of $10^{-18}$.

## 1 Introduction

Over the last two decades, advancements in laser cooling of neutral atoms and ions, laser frequency stabilization, and optical frequency comb techniques have driven significant progress in atomic optical clocks. These developments have enabled clock systematic frequency shift uncertainty to reach the low level of $10^{-18}$, or even lower[1-8], marking the dawn of a new era in metrology of frequency and time. Although the second in the International System of Units (SI) is currently defined by the $^{133}$Cs fountain clock, optical clocks surpass their microwave counterparts by up to two orders of magnitude in performance, positioning them as the leading candidate for redefining the SI second. Moreover, the superior performance of optical clocks has been exploited for relativistic geodesy and fundamental physics experiments such as the search for time-variations of fundamental constants[9-16].

The development of high-precision, robust and compact transportable optical clock (TOC) represents a technological milestone in transitioning from proof-of-principle demonstrations to field applications. Significant progress has been achieved in both ion-trap-based systems (single ion clock) and optical-lattice-based platforms (neutral atoms clock), and various demonstration based on these TOCs are also being carried out extensively[14-20]. Among various neutral-atoms and ions, the $^{171}$Yb$^+$ ion stands out as an exceptional candidate for developing TOCs[19-21]. Firstly, it has two highly favorable optical clock transitions: the electric quadrupole transition of $^2S_{1/2} \to {}^2D_{3/2}$ at 436 nm and the electric octupole transition of $^2S_{1/2} \to {}^2F_{7/2}$ at 467 nm. The 436 nm transition can be used for an *in situ* analysis of the environmental electromagnetic field for its relative higher sensitivity, which provides distinct advantages for systematic evaluations of 467 nm transition. Moreover, both of these two transitions have currently become the secondary representations of the SI second[22]. Secondly, these two transitions exhibit magnetic field insensitivity as they correspond to $m_F = 0 \to m_F = 0$ configurations. Remarkably, the 467 nm transition demonstrates an extraordinary clock state lifetime of 1.6 years[23]. These characteristics provide critical conditions for improving the frequency stability in optical clocks comparison by using correlation spectroscopy technique in the future[24-25]. Thirdly, the $^{171}$Yb$^+$ ion possesses a relatively large atomic mass and the wavelength of its cooling laser is very close to the dissociation wavelength of YbH$^+$ ions[26], which are formed through reactions with residual background gases in the vacuum. These properties enable a significantly extended ion storage time, thereby improving the uptime of the clock[27]. Fourthly, the lasers required for manipulating the $^{171}$Yb$^+$ ion are readily available and compact in design by semiconductor technologies.

Currently, research on $^{171}$Yb$^+$ optical clock has achieved significant progress internationally and the relative systematic uncertainty of 467 nm transition has reached a remarkable level of $2\times10^{-18}$[2-3]. Based on the 436 nm transition of $^{171}$Yb$^+$, a TOC developed within a pilot project for quantum technology in Germany was demonstrated the instability and uncertainty budget at $10^{-17}$ level[15], and a similar prototype was also been developed in Russia[19]. These prototypes of TOC were packaged into standard 19-inch racks and demonstrated promising user operability in initial testing. At present, in China, both the National Institute of Metrology and the Innovation Academy for Precision Measurement Science and Technology of the Chinese Academy of Sciences are performing research on $^{171}$Yb$^+$ optical clocks[28-29]. The latter has a long-term accumulation in developing TOCs and exploring their various applications, laying a good foundation for the development of high-precision and compact TOC using a single $^{171}$Yb$^+$ ion.

This article presents our development of a TOC based on the 467 nm transition of $^{171}$Yb$^+$. A modular trapping-ion system has been developed that uses a linear Paul trap to confine a single $^{171}$Yb$^+$ ion. After optimizing the laser cooling parameters and compensating for the excess micromotion, we performed a spectral analysis of the 467 nm transition. With a Rabi interrogation time of 400 ms, a Fourier limit linewidth about 2.3 Hz was achieved. A long-term closed-loop operation of this clock was performed with a probe time of 180 ms and the frequency instability was evaluated to be $2.2\times10^{-15}/\sqrt{\tau/s}$ by self-comparison, and it has been corrected to represent the instability of a single servo. This performance means that the resolution of frequency measurement can reach the high parts in $10^{-18}$ level with an averaging time of only one day, which provides a good beginning for future improvements.

## 2 Development of the optical clock

### 2.1 Modular and integrated design

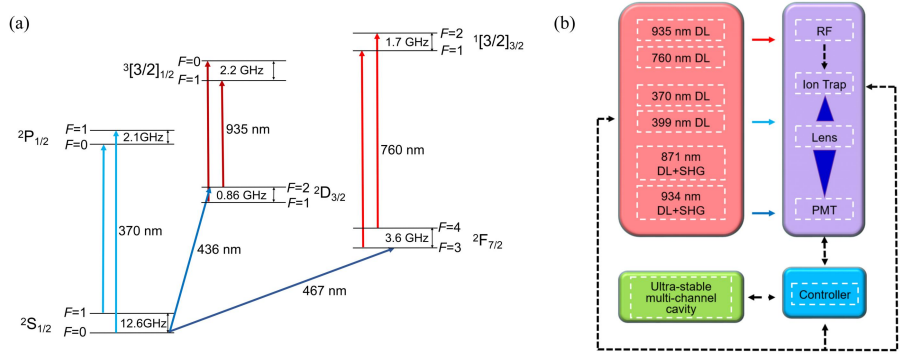

Figure 1. (a) Partial level scheme of $^{171}$Yb$^+$. The colored solid lines with arrows represent the lasers used in the experiment. (b) Schematic diagram of the transportable optical clock.

Fig. 1(a) presents the related atomic level structure of $^{171}$Yb$^+$, delineating the wavelengths utilized for cooling, repumping, and probing the clock transition. Due to the nuclear spin of I = 1/2, hyperfine splitting can be observed across all energy levels. The electric dipole transition from $^2S_{1/2}(F = 1)$ to $^2P_{1/2}(F = 0)$ at 370 nm features a natural linewidth of approximately 19.6 MHz[30], which is optimal for Doppler cooling and fluorescence detection of the ion. Notably, ions in the $^2P_{1/2}$ state have a roughly 0.5%[31] probability of decaying to the metastable $^2D_{3/2}$ state, which possesses a natural lifetime of approximately 53 ms[32]. To maintain the cooling loop, a 935 nm repumping laser is employed. Its frequency is modulated by an electro-optic modulator (EOM) to generate sidebands, which is utilized to retrieve populations from the $^2D_{3/2}$ state. However, due to non-resonant excitation effects, ions in the $^2S_{1/2}(F = 1)$ state may be excited to the $^2P_{1/2}(F = 1)$ state and subsequently undergo spontaneous decay to the ground state of $^2S_{1/2}(F = 0)$. To counteract this effect, a repumper sideband laser with a frequency offset of 14.75 GHz relative to the Doppler cooling laser is employed to generate frequency sidebands by an EOM. This arrangement ensures the efficient removal of ions from the $^2S_{1/2}(F = 0)$ ground state, thus sustaining the cooling loop. Similarly, a secondary repumping laser component shifted by 3.06 GHz is also integrated into the repumping system. Furthermore, collisions with background gas may transfer ions to the $^2F_{7/2}$ state, which has an exceptionally long natural lifetime of up to 1.6 years[23]. To address this problem, a quenching laser at 760 nm is used to reintegrate the ions into the cooling loop. The 467nm laser is used to probe the electric octupole clock transition, while the 436 nm laser is employed to assist in the precise and *in situ* assessment of partial systematic shifts.

The overall design scheme of the TOC is shown in Fig. 1(b). The physical system comprises two parts: the laser stabilization system and the ion trapping system. These modules are connected via optical fibers and cables with each other, ensuring better reliability and flexibility, which are essential for daily operation and maintenance after transportation. For laser stabilization, a compact multi-channel Fabry-Pérot (FP) cavity is employed for frequency stabilization of all the lasers. This strategy not only improves the efficiency of resource utilization in the system but also promotes the development of a compact and reliable transportable $^{171}Yb^+$ optical clock. For ion trapping, the vacuum system of ion trap and surrounding components for radio frequency source and lasers have been specifically redesigned for integration and transportation. All the optical components used for laser transmission are installed through cage-structured modules and fixed near the windows of the vacuum chamber to enhance the spatial stability of laser propagation. In addition, all the optical components used for laser switching and frequency modulation have been modularized to meet the requirement of portability. It is expected that in the subsequent research, these modules will be further integrated, with the ultimate goal of achieving a fully transportable prototype.

**2.2 Laser stabilization system**

All the lasers in Fig. 1(b) are directly generated by semiconductor laser diodes, except for the 436 nm and 467 nm lasers which require the assistance of second harmonic frequency generation. Taking the 467 nm laser as an example, the 934 nm seed laser is first power amplified through a tapered amplifier. The fiber-coupled output with an optical power of approximately 300 mW is fed into a periodically poled lithium niobate (PPLN) crystal for single-pass frequency doubling, and finally, a 467nm laser output of approximately 30 mW is obtained. As for the 934 nm laser transmitted from the crystal, part of it is further utilized, such as being delivered to the reference cavity for frequency stabilization and to the femtosecond optical comb for frequency measurement.

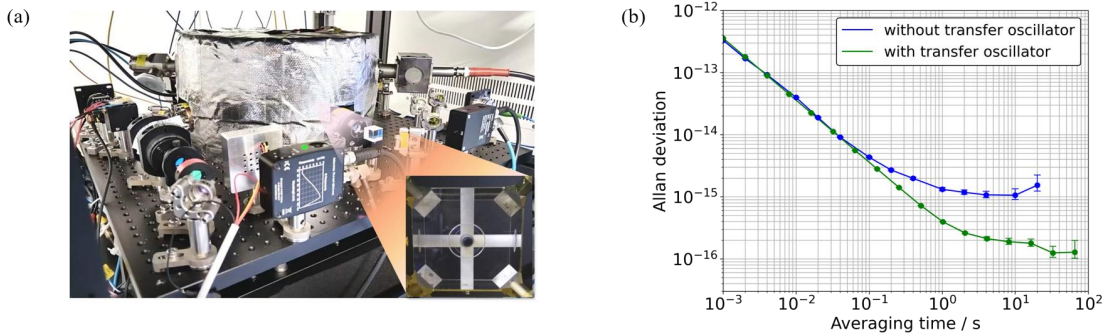

Figure 2. Laser stabilization system and the frequency instability of the clock laser. (a) Physical device of the laser stabilization system with an illustration of the ULE cavity in the lower right corner. (b) The frequency instabilities of the 934 nm laser without (blue line) and with (green line) the additional assistance of transfer oscillator for better performance. The green line was obtained by measuring the instability contributed from the transfer oscillator link (using the beat signal of the 934 nm laser and the optical comb), and then combining it with the instability of the 1064 nm ultra-stable laser.

A compact multi-channel FP cavity made of ultra-low expansion (ULE) glass is used for frequency stabilization of all the lasers. This cubic cavity with a spacer length of 75 mm is installed in a rigid cage to adapt to the application after transportation and is configured with two orthogonal horizontal channels[33-34]. The first channel is optically contacted with a pair of high-reflectivity mirrors (finesse ~ 180,000) for the narrow linewidth laser and the thermal-noise-limited of it is estimated to be about $1\times10^{-15}$. The second channel is optically contacted with a pair of broadband reflector for other lasers. To achieve the lowest laser frequency drift, the cavity is packaged with two layers of gold-plated thermal shielding. Then, all of them are sealed in an ultra-high vacuum chamber with a pressure of $1\times10^{-6}$ Pa, and the temperature is controlled near the point where the coefficient of thermal expansion of ULE is zero. The clock laser is referenced to the cavity with the method of Pound-Drever-Hall frequency stabilization, and the remaining lasers are stabilized using the lock-in technique to simplify the overall system. As shown in Fig. 2(a), both the vacuum system and the optical components for frequency stabilization are mounted on an active vibration isolation platform to minimize the vibration noise. This design not only meets the need for laser frequency stabilization of multi-laser in the $^{171}$Yb$^+$ optical clock system, but also maximizes the utilization of key hardware resources, such as vacuum system, temperature controller and vibration isolation platform. Finally, all the devices are packaged in a cubic aluminum box of dimensions $50 \times 50 \times 35$ cm$^3$ and this box is wrapped with thermal insulation foam to reduce the impact of temperature fluctuation in the environment.

Since only one set of laser frequency stabilization system mentioned above has been established so far, we evaluate the performance of the 934 nm laser by means of a fiber femtosecond comb and a set of 1064 nm ultra-stable laser. Thanks to the adoption of a 30 cm long ULE cavity, the short-term frequency instability of this 1064 nm laser is as low as $1\times10^{-16}$[35], which makes it an excellent reference for evaluation of the 934 nm lasers. As shown in Fig. 2(b), the frequency instability is better than $2\times10^{-15}$ when the averaging time is between 1 s and 20 s after subtracting the linear frequency drift, and it reaches the optimal state at about 10 s. In fact, the short-term stability of the 1064 nm laser can also be transferred by the comb to the 934 nm laser using a transfer oscillator scheme[36] and the improved frequency stability is also presented in Fig. 2(b).

**2.3 Ion trapping system**

The ion trapping system, shown in Fig. 3(a), consists of a compact vacuum chamber with a main body of octagon which is kept below $1\times10^{-8}$ Pa by an ion pump and a getter pump. A linear Paul trap[29], located in the center of the octagon, is employed for our experiment. This trap includes four diagonal blade-shaped electrodes with a spacing of $2r_0 \approx 1.0$ mm, and end-cap electrodes with a spacing of $2z_0 \approx 6.0$ mm. The operation of the trap is facilitated by a helical resonator with resonant radio frequency (rf) of $\Omega = 2\pi \times 26.8$ MHz and a quality factor $Q \approx 330$. Utilizing a rf driving power of roughly 2 W for blade-shaped electrodes and a dc voltage of 200 V for end-caps, the secular motion frequencies of a single $^{171}$Yb$^+$ ion are recorded as $\{f_x, f_y, f_z\} = 2\pi \times \{1.30, 1.32, 0.62\}$ MHz. An imaging lens with an optical aperture of 2 inches and a numerical aperture of 0.38 is employed for the collection of the 370 nm fluorescence emitted by the ion. This fluorescence signal is analyzed using an electron-multiplying CCD (EMCCD) for ion imaging and a photomultiplier tube (PMT) for fluorescence counting. The entire optical configuration around the ion trap employs a modular cage system, ensuring alignment accuracy while providing stablity for building a scalable optomechanical platform. Additionally, a cylindrical magnetic shielding with a diameter of 40 cm and a height of 70 cm encloses the entire system to ensure a stable low magnetic field required during the clock interrogation stage.

Ytterbium comprises multiple natural isotopes, among which the natural abundance of $^{171}$Yb is approximately 14%. To reduce the contamination from non-171 mass isotopes when using unpurified metallic sources, a selective ionization scheme based on two-photon resonant excitation is implemented. Under the ion trap, a ytterbium atomic beam is generated from the oven by resistive heating and then guided to the trap center. The atom is first resonantly excited by a 399 nm laser tuned to the $^1S_0 \rightarrow {}^1P_1$ transition of $^{171}$Yb, and then ionized to $^{171}$Yb$^+$ ion and confined in the ion trap. This selective ionization approach ensures that a single $^{171}$Yb$^+$ ion is loaded within an average time of five minutes.

In the laser cooling experiment, two electro-optic modulators (EOMs) are used to generate 14.75 GHz sidebands for the 370 nm cooling laser and 3.06 GHz sidebands for the 935 nm repumping laser, respectively. Subsequently, laser beams are combined together and then incident on the ion. At the center of the ion trap, their beam waists are approximately 60 μm and 100 μm respectively, and the corresponding laser powers are 3 μW and 300 μW respectively. Similarly, the beams of the 760 nm quenching laser and the 467 nm clock laser are combined together and then incident on the ion. To optimize the fluorescence counts and laser cooling efficiency of a single $^{171}$Yb$^+$ ion, a bias magnetic field of approximately 6 Gauss is applied during the cooling and detection stages. The magnetic field should ideally form a 54.7° angle with the cooling laser's polarization direction[37]. In this experiment, the magnetic field was oriented at 45° relative to the propagation direction of the 370 nm laser. The initial laser polarization was set to horizontal, followed by polarization optimization to maximize fluorescence counts. As shown in Fig. 3(b), these conditions above enable the on-resonance fluorescence counting rate of a single ion to reach 27,000/s, while the typical background counting rate is less than 1,300/s, corresponding to a signal-to-background ratio more than 20. Using the Lorentz approximation, the linewidth is approximately 23.4 MHz.

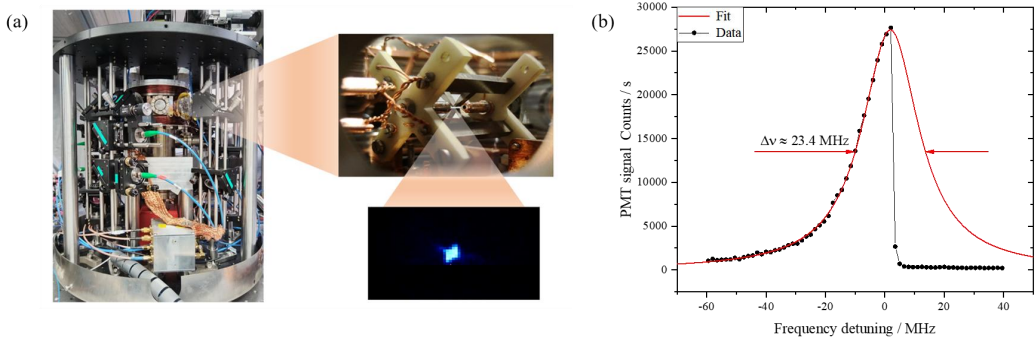

Figure 3. Ion trapping system and the signal of a single $^{171}$Yb$^+$ ion. (a) Physical device of the ion trapping system with a magnified view of the linear blade trap and a single-ion imaging from the EMCCD. (b) Variation of ion fluorescence counts with the frequency detuning of the 370 nm laser, achieving a signal-to-background ratio more than 20. The red solid line is the Lorentz approximation of spectral line.

### 3 Operation of the optical clock

#### 3.1 Interrogation of the clock transition

The interrogation of the electric octupole transition is carried out by use of the so-called electron shelving technique[38]. The relevant time sequence is shown in Fig. 4 and is divided into four stages: laser cooling, state preparation, clock interrogation and fluorescence detection. During the stage of laser cooling, all cooling and repumping lasers are turned on to maintain the laser cooling loop, which lasts for 8 ms. In the second stage, the EOM operation at 14.75 GHz needs to be turned off to enable the ion to undergo a non-resonant transition to the $^2P_{1/2}(F = 1)$ and then spontaneously radiate to the ground state $^2S_{1/2}(F = 0)$, ultimately achieving the selective state preparation. In the clock interrogation stage, all cooling and repumping lasers are turned off, and then the clock laser is turned on to excite the electric octupole transition. It is worth mentioning that in order to maximize the transition probability, the direction of the magnetic field should be adjusted in advance so that the angle between it and the direction of the 467 nm laser is close to 31°, and the laser polarization is horizontally polarized, with an angle of 0° relative to the plane formed by the laser propagation direction and the magnetic field direction[39]. In the fluorescence detection stage, the clock laser is first turned off, then the cooling laser and the repumping laser are turned on to collect the fluorescence in order to determine whether the clock transition has occurred. By scanning the frequency of the clock laser and repeating the above time sequence 20 times at each frequency point, the excitation spectrum of the 467 nm transition can be obtained.

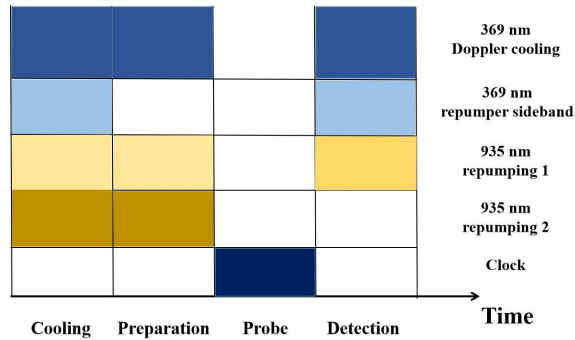

Figure 4. Time sequence for the interrogation of the clock transition. The durations for cooling, preparation, and detection are 8 ms, 0.5 ms, and 4 ms, respectively. The duration of the clock interrogation is variable according to the experimental requirements.

The 467 nm laser was frequency-shifted by acousto-optic modulator (AOM) to approach the central frequency of the clock transition. By choosing the appropriate laser power and scanning step, the excitation spectrum can be obtained at a setting interrogation time. Given the notably weak oscillator strength of the electric octupole transition, an intensity of approximately $10^6$ W/m$^2$ is requisite for achieving a π-pulse excitation within a duration of 100 ms[40]. Using the clock laser only referenced to the 75 mm ultra-stable cavity, a 467 nm spectrum under 100 ms Rabi interrogation time is illustrated in Fig. 5(a). This spectrum exhibits a full width at half maximum (FWHM) of 9.5 Hz with a maximum excitation probability of approximately 0.8. This linewidth has not yet reached the Fourier limit, and it is accompanied by a decrease in transition probability, which is mainly limited by the short-term frequency stability of the clock laser. To verify this issue, we also further used the 934 nm laser after improving the short-term frequency stability through transfer oscillator scheme to attempt a longer clock probing time. This temporary improvement produced an immediate effect and the result is shown in Fig. 5(b). When the clock interrogation time is extended to 400 ms, the maximum transition probability can still be maintained at 0.8 and the corresponding FWHM is narrowed to 2.3 Hz, which is very close to the Fourier limit (2.0 Hz).

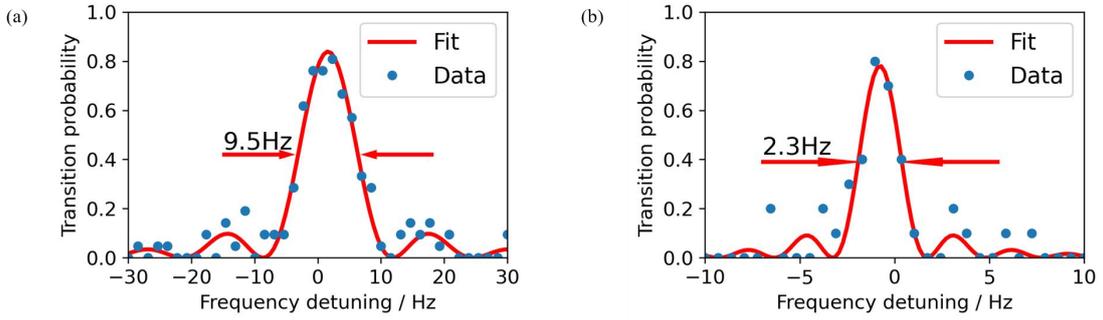

Figure 5. Rabi excitation spectrum of the 467 nm clock transition. Every blue data point is the average over 20 measurement cycles and every red solid line represents a fit to a Rabi line shape. (a) Spectrum before using the transfer oscillator method and the probe time is 100 ms. (b) Spectrum after using the transfer oscillator method and the probe time is 400 ms.

**3.2 Closed-loop locking of the clock**

Unlike the clock with "four-point locking" scheme[41] that usually selects a pair of spectrum as reference, the $^{171}$Yb$^+$ optical clock adopts the "two-point locking" scheme because it only selects one spectrum with Δm$_F$ = 0 as the reference[42]. High operating uptime and excellent frequency stability are crucial for the various applications with transportable optical clocks. To preliminarily evaluate the overall performance of this optical clock prototype, we conducted a nearly continuous locking experiment that lasted for 6 days. Although the electronic controlling system of the prototype has not yet been developed for automation, it has shown good potential for high operating uptime in unattended scenarios from the results in Fig. 6(a). Currently, the 100-meter-long fiber conectted to the optical comb is sometimes severely disturbed, thereby preventing the realization of the fiber noise cancellation. When this occurs, it can only be restored manually after we return to the laboratory. This led to two large gaps in Fig. 6(a), corresponding to data loss due to the problems of the long fiber. Apart from these instances, this prototype remained operation almost all the time, demonstrating a good performance. Overall, the uptime of this clock during this experiment is about 85%, which is a promising start for the first prototype of $^{171}$Yb$^+$ optical clock based on the octupole transition.

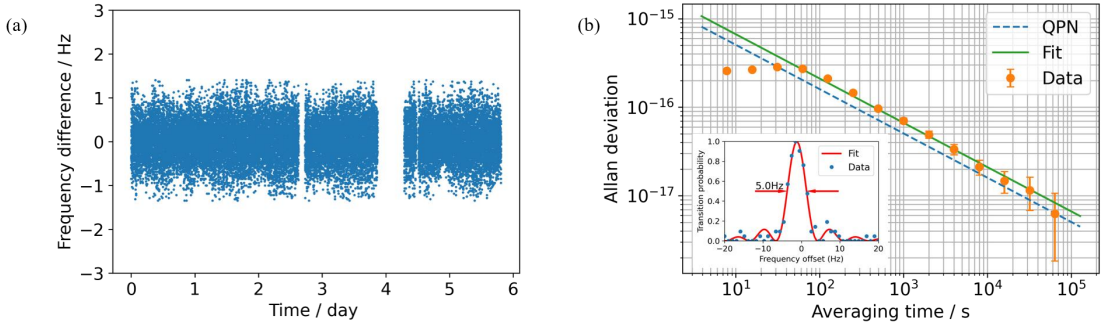

Figure 6. Experimental results of closed-loop locking of the single $^{171}$Yb$^+$ ion optical clock. (a) Uptime of the clock reached 85% over 6-day operation. (b) Frequency instability of clock with the method of self-comparison. The inset illustrates the 180 ms spectrum utilized for clock operation. This clock achieved a frequency instability of $2.2\times10^{-15}/\sqrt{\tau/s}$ through 440,000 seconds of measurement, and it has been corrected to represent the instability of a single servo.

To test the potential of the frequency stability of the clock during the long-term operation mentioned above, a clock interrogation time of 180 ms is adopted and the corresponding spectrum linewidth is about 5 Hz. During the clock locking phase, the inherent light shift induced by the 467 nm laser itself must be accounted for. The light shift in the current system has been measured by interleaved locking under two different probe times (with the corresponding clock laser power ratio of 1:4.1). The estimated light shift with an interrogation time of 180 ms is approximately 13 Hz. The clock laser power incident on the ion is stabilized with the detection of the laser beam transmitted from the vacuum system. The long-term instability of laser power is maintained at the level of $1\times10^{-4}$, corresponding to a fractional frequency instability at the low level of $10^{-18}$. This is negligible for the current frequency instability of the clock. The method of self-comparison[43] is also employed since only one setup of clock is available currently. Fig. 6(b) shows the result of Allan deviation using data sets from Fig. 6(a), and the fractional frequency instability of a single servo is calculated to be $2.2\times10^{-15}/\sqrt{\tau/s}$. Considering that the quantum projection noise limit[44] under the above experimental parameters is $1.6\times10^{-15}/\sqrt{\tau/s}$, it proves that the experimental result is close to the theoretical value. Meanwhile, this result implies that with this prototype, it only takes an averaging time of one day for its precision of frequency measurement to reach the level of $10^{-18}$. Although the current result is achieved with the assistance of a 30 cm ultra-stable cavity, it implies that in future practical applications, if necessary, an independent ultra-stable cavity with much better performance can be added to it as an upgrade accessory, and the frequency stability of the clock can be further enhanced with this flexible and convenient solution.

## 4 Conclusion

In conclusion, this article introduces the recent development towards a TOC based on the 467 nm transition of a single $^{171}$Yb$^+$ ion. Currently, both the ion trapping system and laser stabilization system have been modular designed. A clock transition with a linewidth of 2.3 Hz was achieved using an ultra-stable 467 nm laser, showing a maximum transition probability of more than 80%. During the approximately 6 days operation of this clock, its uptime reached 85% and frequency instability measured by self-comparison was $2.2\times10^{-15}/\sqrt{\tau/s}$. Next, we will further optimize and integrate this prototype, and perform a comprehensive systematic evaluation. We expect to develop a high-precision and compact prototype of TOC with a systematic uncertainty at the low level of $10^{-18}$. This prototype will participate in frequency comparison experiments involving various kinds of optical clocks in China, as well as the field applications for relativistic geodesy.

We thank Dr. R. M. Godun and Dr. N. Huntemann for useful discussions. This work is supported by the National Natural Science Foundation of China (Grant No. 12574291, U21A20431) and the Science and Technology Department of Hubei Province (Grant No. 2025AFA004).